\documentclass[aps,prd,superscriptaddress,twocolumn,amsfonts,floatfix,showpacs]{revtex4}

\usepackage{graphicx}
\usepackage{amsmath}
\def\prl{Phys. Rev. Lett.}
\def\prd{Phys. Rev. D}
\def\cqg{Class. Quantum Grav.}

\def\CovDev{D}
\def\Mpunc{\mathcal{M}}

\begin{document}
   
\title{Trumpet-puncture initial data for black holes}

\author{Jason D. Immerman}
\affiliation{Department of Physics and Astronomy, Bowdoin College,
  Brunswick, ME 04011}

\author{Thomas W. Baumgarte}
\altaffiliation{Also at Department of Physics, University of Illinois at
  Urbana-Champaign, Urbana, IL 61801}
\affiliation{Department of Physics and Astronomy, Bowdoin College,
  Brunswick, ME 04011}
  
\begin{abstract}
We propose a new approach, based on the puncture method, to construct black hole initial data in the so-called trumpet geometry, i.e.~on slices that asymptote to a limiting surface of non-zero areal radius.  Our approach is easy to implement numerically and, at least for non-spinning black holes, does not require any internal boundary conditions.  We present numerical results, obtained with a uniform-grid finite-difference code, for boosted black holes and binary black holes.  We also comment on generalizations of this method for spinning black holes.
\end{abstract}


\maketitle

Numerical simulations of black hole spacetimes have recently experienced a dramatic breakthrough (see \cite{Pre05b,BakCCKM06a,CamLMZ06} as well as numerous later publications).  Most of these simulations now adopt some variation of the BSSN formulation \cite{ShiN95,BauS99} together with the moving puncture \cite{BakCCKM06a,CamLMZ06} method to handle the black hole singularities.  

The {\em moving-puncture} method is based on a set of empirically found coordinate conditions, namely the ``1+log" slicing condition for the lapse \cite{BonMSS95} and a ``$\bar \Gamma$-freezing" gauge condition for the shift \cite{AlcBDKPST03}.  As demonstrated by \cite{HanHPBO06,HanHOBGS06,Bro08,HanHOBO08}, dynamical simulations of a Schwarzschild spacetime
using these coordinate conditions settle down to a spatial slice that terminates at a non-zero areal radius, and hence does not encounter the spacetime singularity at the center of the black hole.  An embedding diagram of such a slice, which suggests the name {\em trumpet} data, is shown in Fig.~2 of \cite{HanHOBO08}.

Typically, moving-puncture simulations adopt initial data that are constructed using the {\em puncture} method \cite{BeiO94,BeiO96,BraB97}.  As we explain in more detail below, the central idea of the puncture method is to write the conformal factor as a sum of an analytically known, singular background term, and a correction term that is unknown but regular.  The equations for the correction term can then be solved everywhere, without any need for excision or any other means of dealing with the black hole singularity.  To date, all applications of this method have adopted Schwarzschild data on a slice of constant Schwarzschild time as the background solution.  These data connect spatial infinity in one universe with spatial infinity in another universe; the resulting initial data therefore represent {\em wormhole} data.

Clearly, it would be desirable to produce initial data that represent black holes as trumpets rather than wormholes, since otherwise moving-puncture evolutions will drive the individual black holes to a trumpet geometry.  Problems with one possible approach, based on a stationary 1+log slicing in the context of the conformal thin-sandwich decomposition, were described in \cite{BauELMOST09}.  Recently, trumpet initial data on hyperboloidal slices using an excision method were constructed in \cite{BucPB09}.  In this paper we demonstrate how such data can be produced by generalizing the puncture method, which does not require any excision or internal boundary conditions, and is very easy to implement numerically.  

In a 3+1 decomposition, Einstein's equations split into a set of evolution equations and two constraint equations, namely the Hamiltonian and the momentum constraints (see, e.g., \cite{Coo00,BauS03} for a review).
Assuming conformal flatness, maximal slicing and vacuum, the Hamiltonian constraint becomes
\begin{equation} \label{eq:ham1}
\bar \CovDev^2 \psi = - \frac{1}{8} \psi^{-7} \bar A_{ij} \bar A^{ij},
\end{equation}
while the momentum constraint decouples from the Hamiltonian constraint and reduces to 
\begin{equation} \label{eq:mom1}
\bar \CovDev_i \bar A^{ij} = 0.
\end{equation}
Here $\psi$ is the conformal factor, $\bar A^{ij}$ is the conformally rescaled traceless part of the extrinsic curvature, $\bar \CovDev_i$ is the covariant derivative associated with the conformally related spatial metric $\bar \gamma_{ij}$, and $\bar \CovDev^2 \equiv \bar \CovDev_i \bar \CovDev^i$ is the Laplace operator.  Analytical solutions to the momentum constraint (\ref{eq:mom1}) are given by the Bowen-York solutions \cite{BowY80};  we will be particularly interested in the boosted black hole solution
\begin{equation} \label{eq:BY}
\bar A^{ij}_{\rm P} = \frac{3}{2 r^2} \left(P^i n^j + P^j n^i - (\bar \gamma^{ij} - n^i n^j) n_k P^k) \right)
\end{equation}
where $P^i$ is the linear momentum and $n^i = x^i/r$ is the spatial normal vector pointing away from the puncture at $r=0$.   Given the linearity of the momentum constraint, spacetimes containing multiple black hole can be constructed using a superposition of several terms of the form (\ref{eq:BY}).  The initial data are then completed by solving the Hamiltonian constraint (\ref{eq:ham1}) for the conformal factor.

The central idea in the puncture method \cite{BeiO94,BeiO96,BraB97} is to write the conformal factor as a sum
\begin{equation} \label{eq:psi_dec}
\psi = \psi_0 + u,
\end{equation}
where the ``background" conformal factor $\psi_0$ is given analytically and absorbs the singular parts of $\psi$, whereas $u$ remains regular and is solved for numerically.  We will similarly write 
\begin{equation} \label{eq:A_dec}
\bar A^{ij} = \bar A^{ij}_0 + \bar A^{ij}_{\rm P},
\end{equation}
where $\bar A^{ij}_{\rm P}$ could be replaced with any other solution to the momentum constraint (\ref{eq:mom1}).

In all applications to date, the background data have been taken to be those of a Schwarzschild black hole on a slice of constant Schwarzschild time,
\begin{equation}
\psi_0 = 1 + \frac{\Mpunc}{2 r}; ~~~~ \bar A^{ij}_0 = 0.
\end{equation}
Here $\Mpunc$ is a free mass parameter.  Evidently, for vanishing momentum $P^i = 0$, the Hamiltonian constraint (\ref{eq:ham1}) is solved identically by $u = 0$.  For non-zero momentum, the Hamiltonian constraint becomes a regular equation for $u$ that can be solved on $\mathbb{R}^3$ without any need for special treatment of the singularity at $r = 0$.   This approach results in {\em wormhole} initial data that, for a single black hole, connect two separate spatial infinities.

To construct {\em trumpet} initial data, we propose to use a trumpet slicing of the Schwarzschild spacetime as background data.  Maximally sliced trumpet data were discussed in \cite{HanHOBGS06}, and an analytic solution, parametrized by the areal radius $R$, is given in \cite{BauN07}.  In particular, the conformal factor is
\begin{eqnarray} \label{eq:psi1}
\psi_0 
&=&
\left[ 
\frac
{4 R} 
{2 R +  \Mpunc + (4 R^2 + 4 \Mpunc R + 3 \Mpunc^2)^{1/2} }
\right]^{1/2} \\
&  & \times 
\left[ 
\frac
{8 R +  6 \Mpunc + 3 ( 8 R^2 + 8 \Mpunc R + 6 \Mpunc^2 )^{1/2} } 
{(4 + 3 \sqrt{2})(2 R - 3 \Mpunc) }
\right]^{1/2\sqrt{2}} 
\nonumber
\end{eqnarray}
with the isotropic radius $r$ given by
\begin{eqnarray} \label{eq:r}
r & = & \left[ 
\frac{2 R +  \Mpunc + (4 R^2 + 4 \Mpunc R + 3 \Mpunc^2)^{1/2} } {4} 
\right]  \\
& &  \times 
\left[ \frac{(4 + 3 \sqrt{2})(2 R - 3 \Mpunc) }
{8 R +  6 \Mpunc + 3 ( 8 R^2 + 8 \Mpunc R + 6 \Mpunc^2 )^{1/2} } 
\right]^{1/\sqrt{2}} . \nonumber
\end{eqnarray}
Asymptotically, $\psi_0$ behaves as
\begin{equation}
\psi_0 = \left\{ \begin{array}{ll}
\displaystyle \left( \frac{3\Mpunc}{2r} \right)^{1/2} ~~~~~ & r \rightarrow 0, \\[3mm]
\displaystyle 1 + \frac{\Mpunc}{2r} &  r \rightarrow \infty.
\end{array} \right.
\end{equation}
The limit surface $r \rightarrow 0$ corresponds to a sphere of areal radius $R = \psi^2 r \rightarrow 3 \Mpunc /2$.  The corresponding background extrinsic curvature is
\begin{equation}
\bar A^{ij}_0 = \frac{3 \sqrt{3} \Mpunc^2}{4 r^3}(\bar \gamma^{ij} - 3 n^i n^j).
\end{equation} 
It is straightforward to verify that the momentum constraint (\ref{eq:mom1}) is satisfied by 
$\bar A^{ij}_0$, and, given its linearity, by any combination of the form (\ref{eq:A_dec}).  The Hamiltonian constraint (\ref{eq:ham1}) now becomes
\begin{equation} \label{eq:ham2}
\bar \CovDev^2 u = - \frac{1}{8} (\psi_0 + u)^{-7} \bar A_{ij} \bar A^{ij} + 
\frac{1}{8} \psi_0^{-7} \bar A^0_{ij} \bar A^{ij}_0,
\end{equation}
where we have used the fact that the background solution must satisfy the Hamiltonian constraint, $\bar \CovDev^2 \psi_0 = - \psi_0^{-7} \bar A^0_{ij} \bar A^{ij}_0 / 8$.   In the following we will focus on black holes carrying linear momentum, and will abbreviate
\begin{equation}
\bar A^2 \equiv
\bar A_{ij} \bar A^{ij} = \bar A^2_0 + 2 \bar A^2_{\rm P0} + \bar A^2_{\rm P}
\end{equation}
with
\begin{subequations}
\begin{eqnarray}
\bar A_0 ^2 & \equiv & \bar A^0_{ij} \bar A^{ij}_0 = \frac{81}{8} \, \frac{\Mpunc^4}{r^6} \\
\bar A_{\rm P0}^2 & \equiv & \bar A^{\rm P}_{ij} \bar A^{ij}_0 = - \frac{ 27 \sqrt{3}}{4} \, \frac{\Mpunc^2 P}{r^5} \, \cos \theta \\
\bar A_{\rm P}^2 & \equiv & 
\bar A^{\rm P}_{ij} \bar A^{ij}_{\rm P} =   \frac{9}{2} \, \frac{P^2}{r^4} \, (1 + 2 \cos^2 \theta).
\end{eqnarray}
\end{subequations}
Here $P$ is the magnitude of $P^i$, and $\theta$ is the angle between $P^i$ and $n^i$ (both with respect to $\bar \gamma_{ij}$). 

Before proceeding it is useful to analyze the properties of Eq.~(\ref{eq:ham2}) and its solutions in the vicinity of the puncture at $r=0$.   Assuming $u \ll \psi_0$, we keep only the leading order terms on the right hand side of Eq.~(\ref{eq:ham2}) to find
\begin{equation} \label{eq:ham3}
\bar \CovDev^2 u \approx \frac{1}{\sqrt{2}} \left( \frac{\Mpunc}{r} \right)^{3/2} \, \frac{P \cos \theta}{\Mpunc^3}
+ \frac{7}{4} \, \frac{u}{r^2}.
\end{equation}
Note that the second term on the right hand side forces all regular solutions to vanish at the puncture $r=0$,
\begin{equation} \label{eq:u_punc_single}
u_{\rm punc} =0.
\end{equation}  
To see this, consider a solution that approaches a non-zero value at $r=0$; the second term on the right hand side would then scale with $r^{-2}$ at $r=0$, this would lead to solutions $u \propto \ln r$, which diverge at the origin.   The general, axisymmetric, regular solution in the vicinity of the origin can be found to be
\begin{eqnarray} \label{eq:u_limit}
u & = & - \frac{1}{3 \sqrt{2}} \frac{P}{\Mpunc} \left( \frac{r}{\Mpunc} \right)^{1/2} \cos \theta \nonumber \\
& & + \sum_{\ell = 0}^\infty C_\ell \, r^{-1/2 + \sqrt{2 + \ell(\ell+1)}} P_\ell(\cos \theta),
\end{eqnarray}
where the coefficients $C_\ell$ are arbitrary (but will be determined by outer boundary conditions), and where the $P_\ell(\cos \theta)$ are Legendre polynomials of order $\ell$.  Close to the puncture the solution is dominated by the first term, which scales with $r^{1/2}$.  In the following we will refer to this first term as the ``limiting solution".

We find numerical solutions to Eq.~(\ref{eq:ham2}) using a modification of the code described in \cite{Bau00}.  The code is a simple uniform-grid, finite-difference, cell-centered code that inverts matrices using {\sc PETSc} software \cite{PETSc97}.  Solutions are constructed iteratively and are accurate to second order in the grid spacing for functions that can be differentiated at least twice; at the puncture, the lack of differentiability of $u$ reduces the order of convergence.   

Asymptotic flatness suggests that at spatial infinity $u$ should approach zero.  As an approximation to this outer boundary condition we impose a Robin condition $\partial_r (r u) = 0$ at the outer boundary of our grid.

Clearly, the accuracy of solutions can be improved in many ways, for example by factoring out the square-root behavior of the solution at the puncture, by using mesh-refinement, by imposing the outer boundaries at larger separation, or by using more efficient numerical techniques.  Our purpose here, however, is to demonstrate that solutions can be found with very simple techniques and without the need of any interior boundary conditions.

Care has to be taken that $u=0$ at the puncture $r=0$ at every step of an iteration, since otherwise the second term on the right hand side of Eq.~(\ref{eq:ham3}) will lead to singular behavior in the next iteration step.  One way to avoid this problem would be to explicitly enforce a boundary condition $u=0$ at $r=0$. However, we have found that this is not necessary if the ``offending" term is absorbed in the operator itself.  In practice, we therefore solve the iteration
\begin{eqnarray} \label{eq:ham_iter}
& & \bar \CovDev^2 u^{N+1} - \frac{7}{8} \psi_0^{-8} \bar A_0^2 u^{N+1} = \\ 
& & ~ - \frac{1}{8} (\psi_0 + u^N)^{-7} \bar A^2   + 
	\frac{1}{8} \psi_0^{-7} \left(1 - \frac{7 u^N}{\psi_0} \right) \bar A_0^2. \nonumber 
\end{eqnarray}
Clearly, when convergence has been achieved with $u^{N+1} = u^N$ to within a desired accuracy, this equation is equivalent to (\ref{eq:ham2}).  We start the iteration (\ref{eq:ham_iter}) either with $u^0=0$ or, when computing sequences of solutions, with the solution $u$ for the previous member of the sequence.

\begin{figure}
\includegraphics[width=3in]{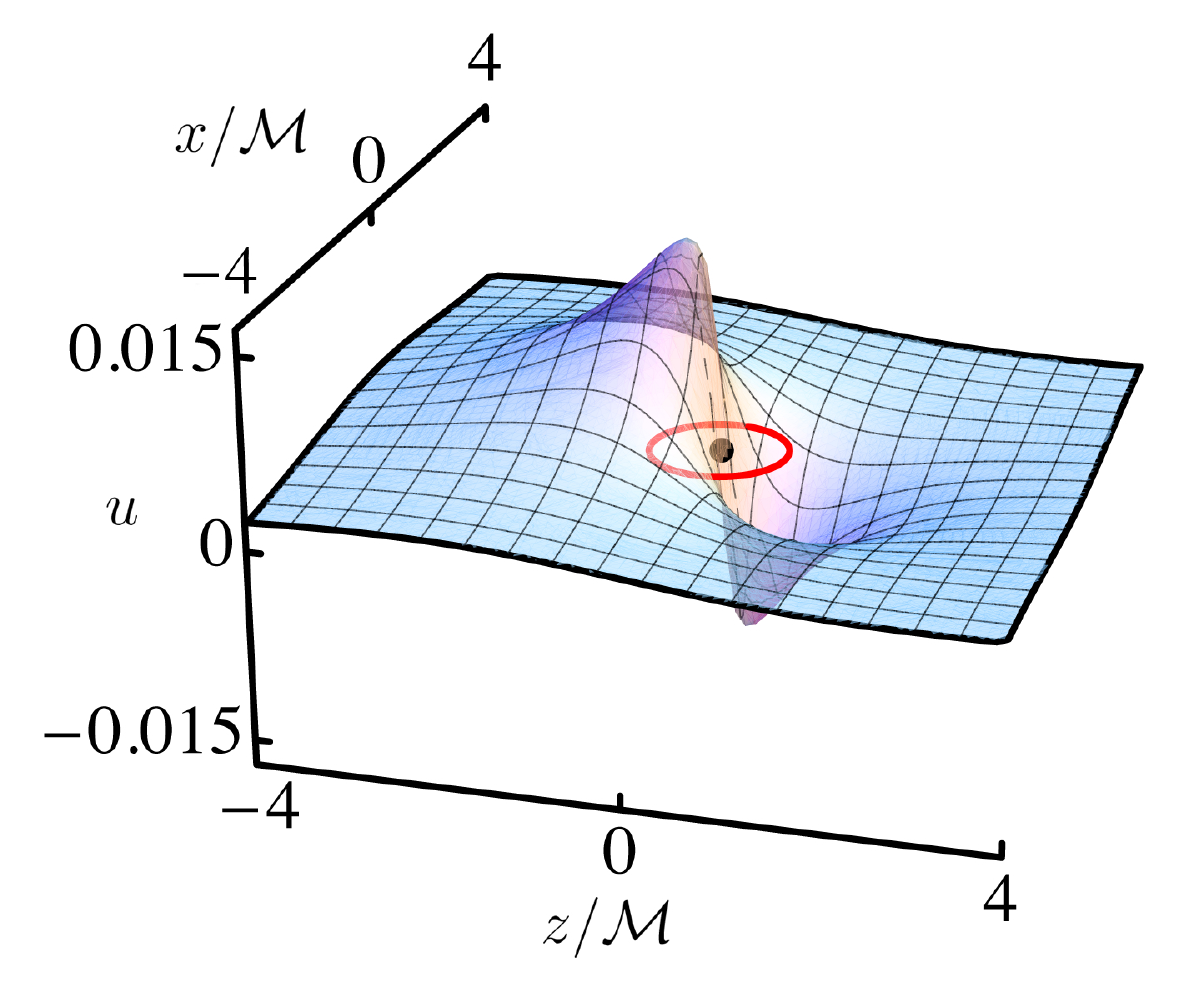}
\caption{A surface plot of $u$ in the $x-z$ plane, for a momentum of magnitude $P = 0.2 \Mpunc$ pointing in the positive $z$ direction.  The black dot marks the location of the puncture, the red circle the location of the apparent horizon.}
\label{Fig1}
\end{figure}

\begin{figure}
\includegraphics[width=3in]{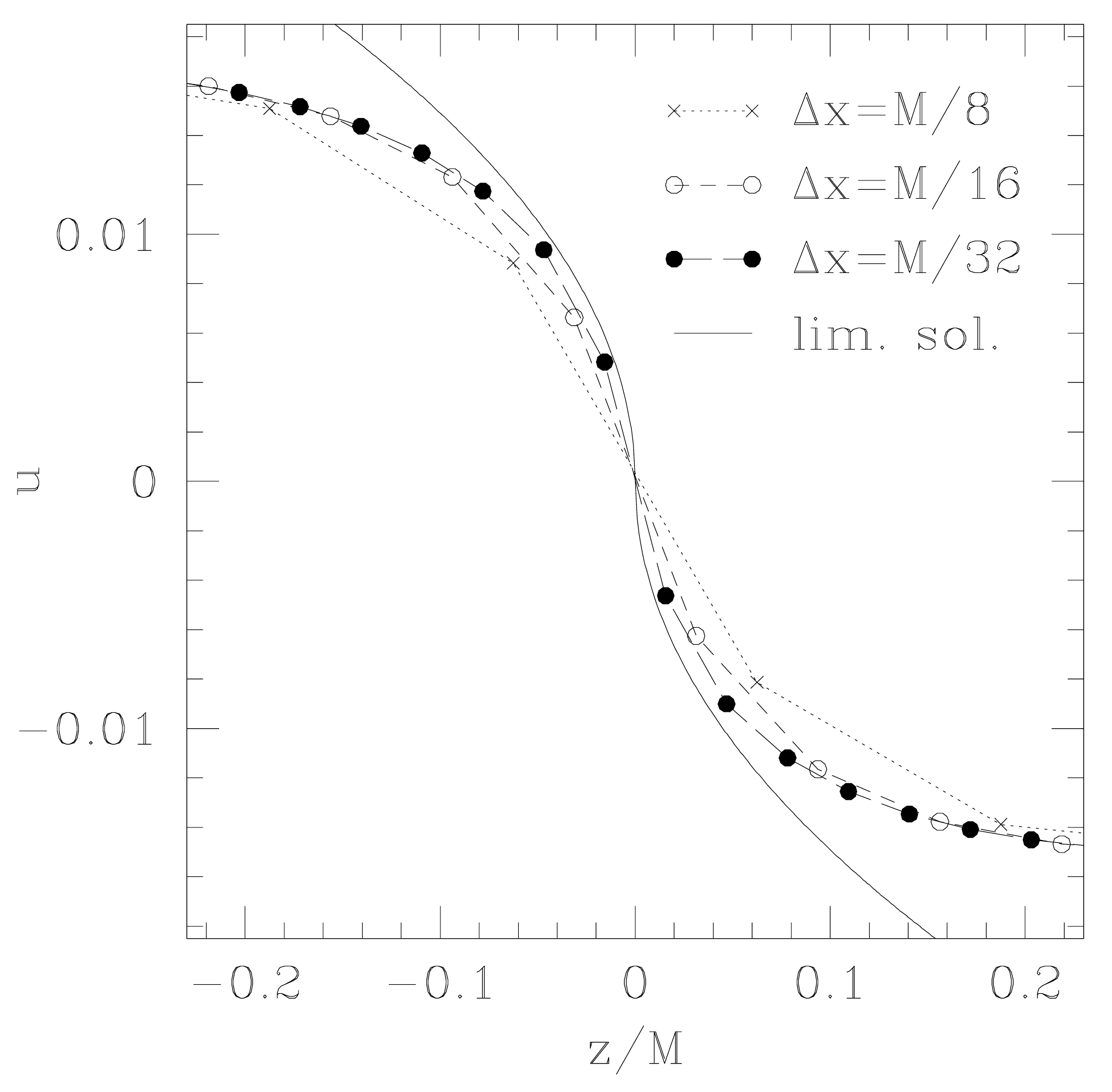}
\caption{Solutions $u$, shown on the $z$-axis, for a momentum $P = 0.2 \Mpunc$ pointing in the positive $z$-direction.  The solutions are shown for three different grid resolutions, and are found to converge to the limiting solution (\ref{eq:u_limit}) in the vicinity of the puncture at $r=0$.}
\label{Fig2}
\end{figure}

\begin{figure}
\includegraphics[width=3in]{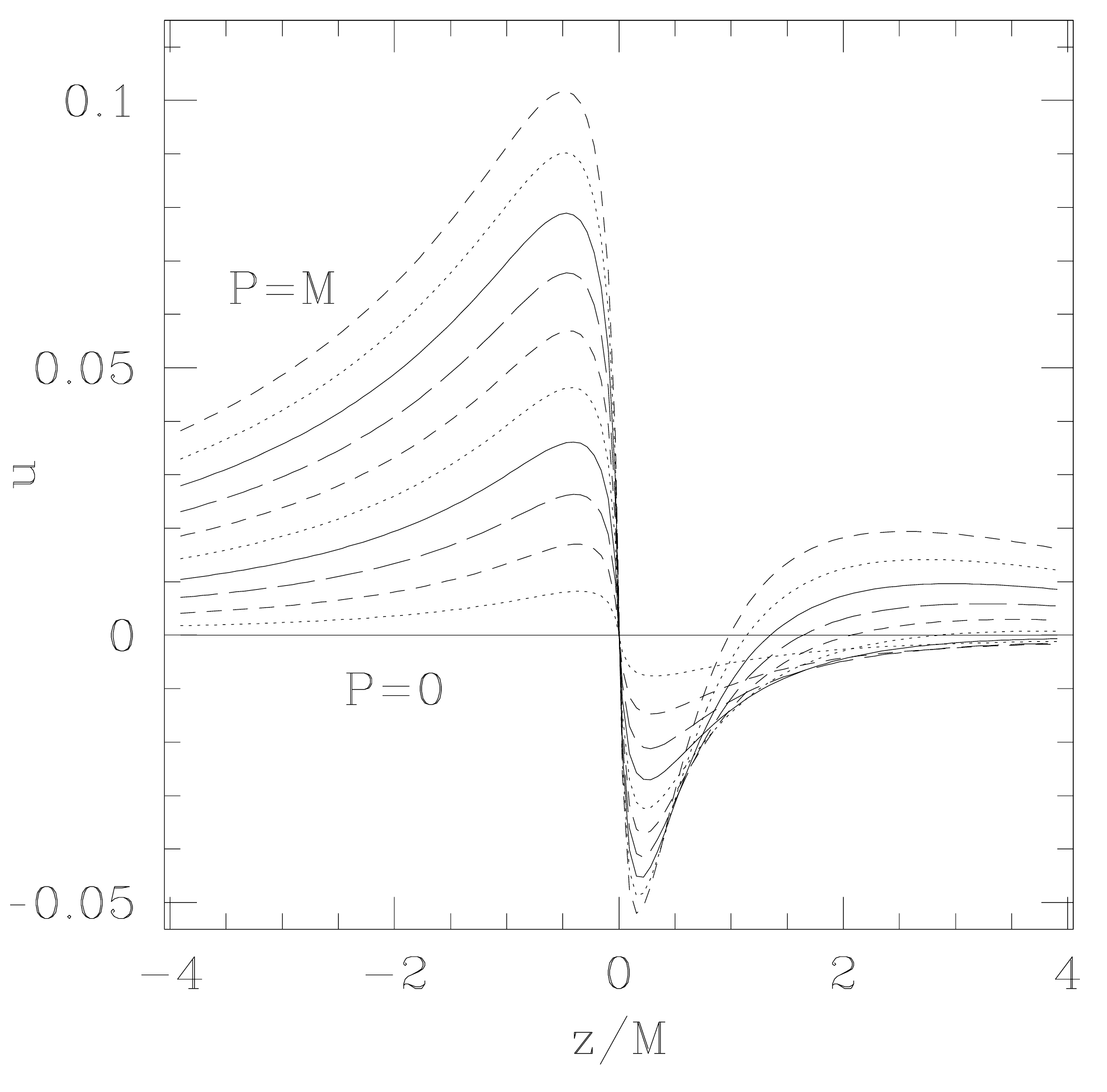}
\caption{A family of solutions $u$, shown on the $z$-axis, for momenta of magnitude $0 \leq P \leq \Mpunc$ pointing in the positive $z$-direction, in increments of $\Delta P = 0.1 \Mpunc$.}
\label{Fig3}
\end{figure}

In Fig.~\ref{Fig1} we show a solution $u$ for a momentum of magnitude $P = 0.2 \Mpunc$ pointing in the positive $z$-direction.  In Fig.~\ref{Fig2} we show results for the same momentum, but projected onto the $z$-axis and obtained for three different grid resolutions $\Delta x = \Mpunc/8$, $\Delta x = \Mpunc/16$, and $\Delta x = \Mpunc/32$.  Here the outer boundaries are imposed at $X_{\rm max} = Y_{\rm max} = Z_{\rm max} = 2 \Mpunc$.  We also include a graph of the limiting solution (\ref{eq:u_limit}); the numerical solutions converge to this limiting solution in the vicinity of the puncture at $r=0$.

\begin{figure}
\includegraphics[width=3in]{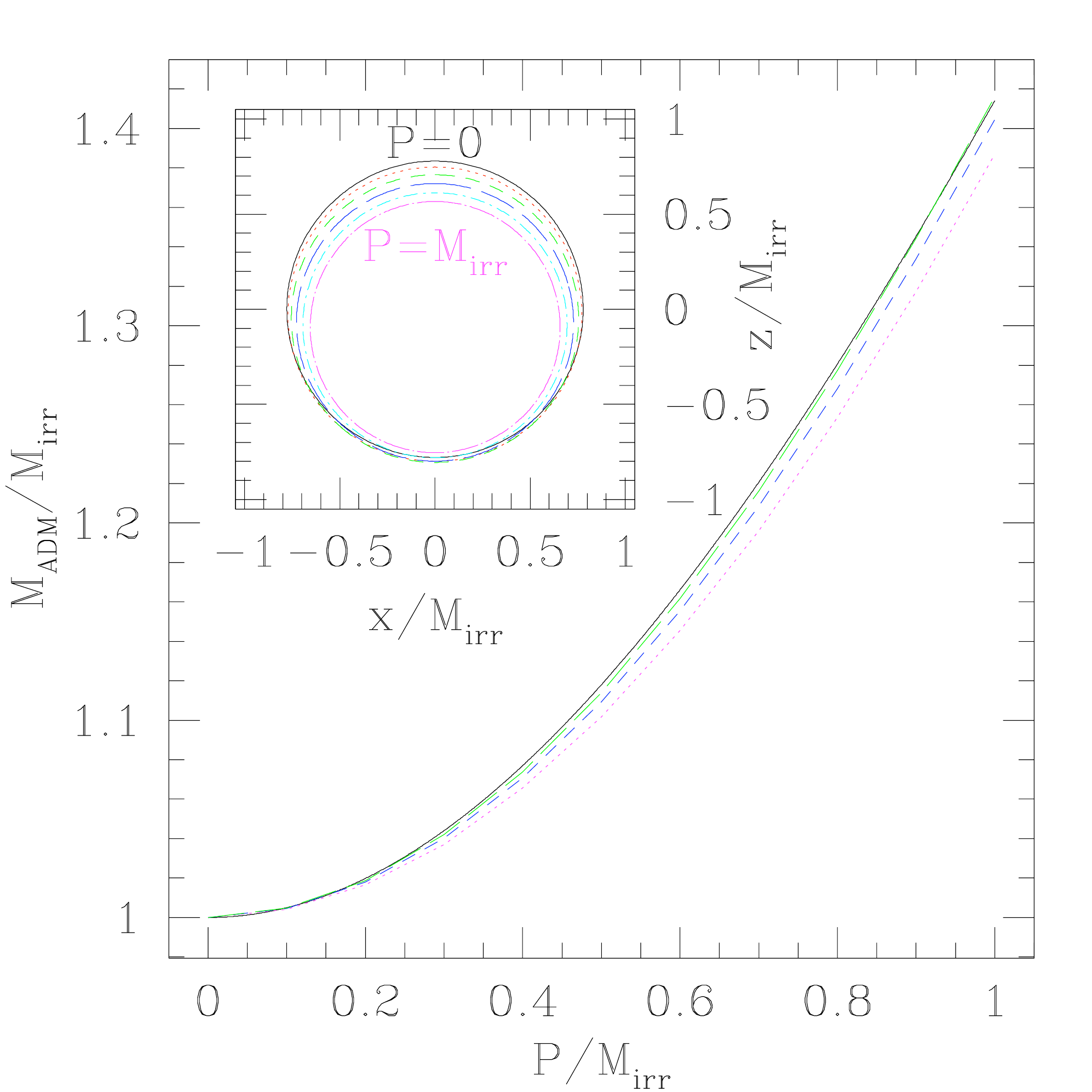}
\caption{The ADM mass $M_{\rm ADM}$ as a function of $P$ for the family of solutions shown in Fig.~\ref{Fig2}.  We graph results for three different locations of the outer boundary, namely at $2 \Mpunc$ (dotted line), $4\Mpunc$ (short dashed line) and $8 \Mpunc$ (long dashed line), at constant grid resolution of $\Delta x = \Mpunc/8$.  The solid line represents the result for a boosted Schwarzschild black hole, $M_{\rm ADM} = \gamma M_{\rm irr}$.  The inset shows the locations of apparent horizons for the same sequence, in increments of $\Delta P = 0.2 \Mpunc$.}
\label{Fig4}
\end{figure}

In Fig.~\ref{Fig3} we show results for $u$ for momenta with magnitudes $0 \leq P \leq 1.0 \Mpunc$, all pointing in the positive $z$-direction.  For these calculations we imposed the outer boundaries at $X_{\rm max} = Y_{\rm max} = Z_{\rm max} = 4 \Mpunc$ and used a resolution of $\Delta x = \Mpunc/16$.  In the inset of Fig.~\ref{Fig4} we show the apparent horizons for this sequence, located with the apparent horizon finder described in \cite{BauCSST96}.  For small values of $P$, the horizons are simply translated in a direction opposite to that of the momentum (compare \cite{CooY90,DenBP06}).  The irreducible mass $M_{\rm irr}$, as approximated by the proper area of the apparent horizon, remains equal to the mass parameter $\Mpunc$ along this sequence, to within the accuracy of our code.  This finding has been confirmed by research performed concurrently with ours \cite{HanHO09}, and differs from the case for wormhole data (see, for example, the analytical treatment of \cite{DenBP06}).  

The Arnowitt-Deser-Misner (ADM) mass of our solutions can be computed from
\begin{eqnarray} \label{eq:adm_mass}
M_{\rm ADM} & = & - \frac{1}{2 \pi} \oint_{\infty} \bar \CovDev^i \psi d \bar S_i \nonumber \\
& = & - \frac{1}{2 \pi} \oint_{\infty} \bar \CovDev^i \psi_0 d \bar S_i - \frac{1}{2 \pi} \int \bar \CovDev^2 u dV
\\
& = &  \Mpunc + \frac{1}{16 \pi} \int \left( (\psi_0 + u)^{-7} \bar A^2 -
\psi_0^{-7} \bar A_0^2 \right) dV. \nonumber
\end{eqnarray}
In Fig.~\ref{Fig4} we graph the ADM mass as a function of $P$, for a given grid resolution of $\Delta x = \Mpunc/8$ and for three different locations of the outer boundaries.  We have also included the result for a boosted Schwarzschild black hole, $M_{\rm ADM} = \gamma M_{\rm irr}$, where $\gamma$ is the gamma factor.

To construct a binary, with black hole punctures located at coordinate locations ${\bf C}_+$ and ${\bf C}_-$, we write the conformal factor $\psi$ as
\begin{equation} \label{eq:psi_binary}
\psi  = \psi_+ + \psi_- - 1+ u,
\end{equation}
where $\psi_+$ and $\psi_-$ are the same as $\psi_0$ in Eq.~(\ref{eq:psi1}), except centered on ${\bf C}_+$ or ${\bf C}_-$.  We subtract a value of unity in Eq.~(\ref{eq:psi_binary}) so that $\psi$ still approaches unity at infinity with $u$ approaching zero.  Similarly, the extrinsic curvature $\bar A^{ij}$ is now given by
\begin{equation} \label{eq;A_binary}
\bar A^{ij} = \bar A^{ij}_{0+} + \bar A^{ij}_{0-} + \bar A^{ij}_{P+} + \bar A^{ij}_{P-}.
\end{equation}
The Hamiltonian constraint then becomes
\begin{equation} \label{eq:ham4}
\bar \CovDev^2 u = - \frac{1}{8} (\psi_+ + \psi_- - 1 + u)^{-7} \bar A^2 
	+ \frac{1}{8} \psi_+^{-7} \bar A^2_{0+} 
	+ \frac{1}{8} \psi_-^{-7} \bar A^2_{0-}.
\end{equation}
Clearly, this approach can be generalized to more black holes in a similar fashion.  Expanding the right hand side of Eq.~(\ref{eq:ham4}) about ${\bf C}_+$, say, to leading order in $r_+$, we find
\begin{equation}
\bar \CovDev^2 u = \frac{1}{\sqrt{2}} \left(\frac{\Mpunc_+}{r_+}\right)^{3/2} \frac{P_+ \cos \theta_+}{\Mpunc_+^3} + \frac{7}{4} \frac{\psi_- - 1 + u}{r_+^2}.
\end{equation}
Repeating the arguments below Eq.~(\ref{eq:u_punc_single}), we now find that  the value of $u$ at a puncture must be
\begin{equation} \label{eq:u_punc_binary}
u_{\rm punc}^+ = \left. \psi_-\right|_{r_+ = 0} - 1,
\end{equation}
and similar for the other puncture.  In the limit of infinite binary separation, this value reduces to $u_{\rm punc}  = 0$ in Eq.~(\ref{eq:u_punc_single}), as expected.

\begin{figure}
\includegraphics[width=3in]{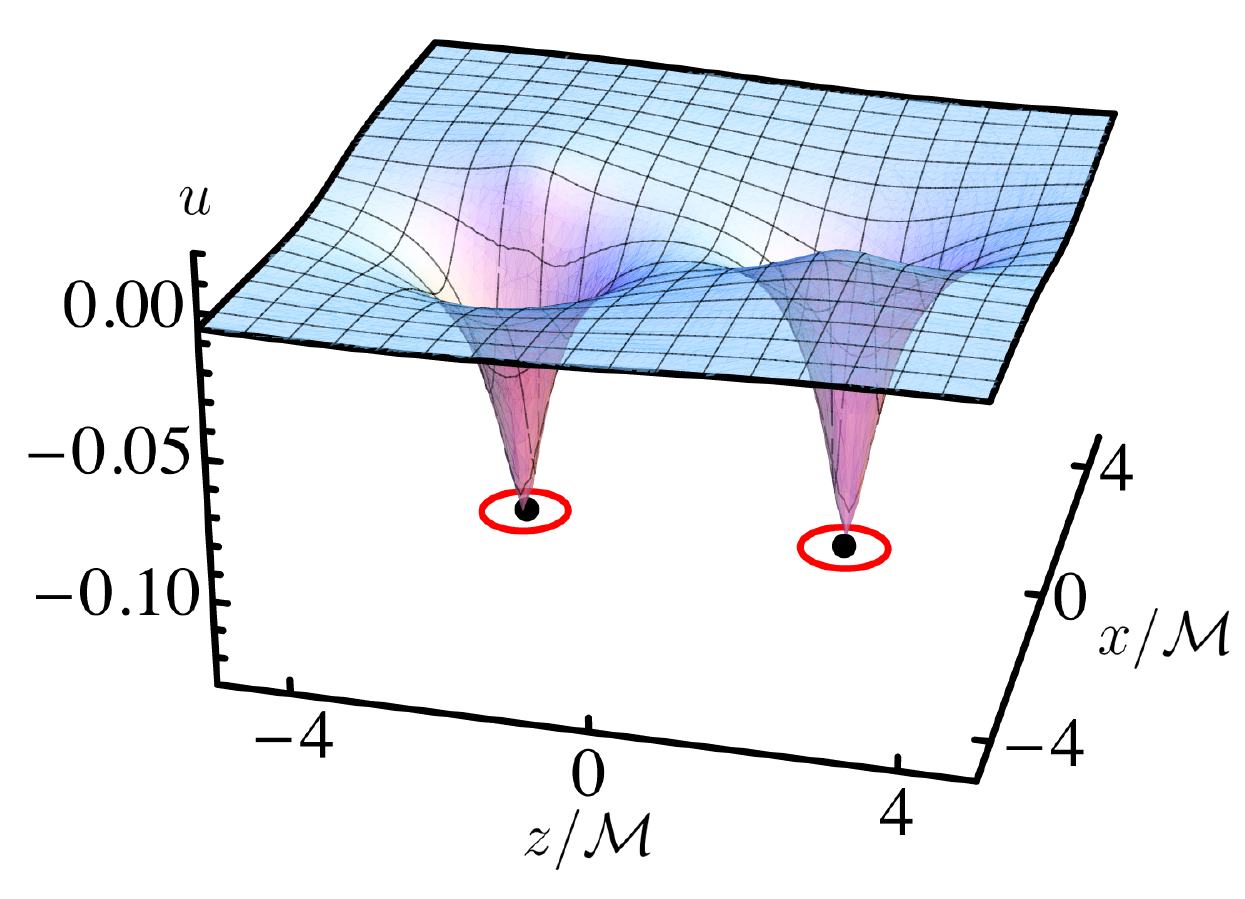}
\caption{A solution $u$ for the binary configuration described in the text.  As in Fig.~\ref{Fig1}, the black dots denote the two punctures, and the red circles the apparent horizons.}
\label{Fig5}
\end{figure}

To demonstrate the feasibility of this approach, we show in Fig.~\ref{Fig5} results for $u$ in the $x-z$ plane for an equal-mass binary with coordinate parameters ${\bf C}_+ = - {\bf C}_- = (0,0,2.25 M_{\rm irr})$ and ${\bf P}_+ = - {\bf P}_- = (59/90M_{\rm irr},0,0)$ (which were identified with the innermost stable circular orbit in \cite{Bau00}).  We expect that similar ``trumpet-puncture" solutions may play a very useful role as initial data  for future dynamical simulations of binary black holes.

Before closing we point out that for spinning black holes our approach has to be modified.  To construct spinning black holes, the Bowen-York solution $\bar A^{ij}_{\rm P}$ in Eq.~(\ref{eq:BY}), which describes black holes carrying a linear momentum, has to replaced with the corresponding expression 
\begin{equation} \label{eq:BY_spin} 
\bar A^{ij}_{\rm S} = \frac{6}{r^3} n^{(i} \bar \epsilon^{j)kl} J_k n_l
\end{equation}
for black holes carrying angular momentum $J_k$.  This expression, however, scales with $r^{-3}$ instead of $r^{-2}$.  As a consequence, the leading-order term in the expansion (\ref{eq:ham3}) then scales with $r^{-5/2}$, meaning that solutions $u_{\rm S}$ now scale with $r^{-1/2}$ instead of $r^{1/2}$.   Noting that $\bar A_{ij}^{\rm S} \bar A^{ij}_{0} = 0$ and 
\begin{equation}
\bar A^2_{\rm S} \equiv \bar A_{ij}^{\rm S} \bar A^{ij}_{S} = \frac{18 J^2}{r^6} \sin^2 \theta
\end{equation}
it can be shown that, in the vicinity of $r=0$ and to leading order in both $r$ and $J^2$, solutions $u_{\rm S}$ with $u_{\rm S} \ll \psi_0$ are 
\begin{equation}
u_{\rm S} = \frac{1}{12} \left( \frac{J}{\Mpunc^2} \right)^2 \left( \frac{2\Mpunc}{3r} \right)^{1/2} 
\left(3 - \cos^2 \theta \right)
\end{equation}
(compare \cite{LovOPC08,BucPB09,DaiG09}).
The corrections $u_{\rm S}$ are no longer regular at the puncture, and scale with $r^{-1/2}$ exactly as the background conformal factor $\psi_0$.  Contrary to linear momentum, the angular momentum of a trumpet black hole does affect the geometry of the limit surface $r \rightarrow 0$.  The angular momentum increases the overall proper area of the limit surface, and also deforms the surface so that it can be represented by an oblate spheriod rather than a sphere.  

Given the singular behavior of $u_{\rm S}$, these solutions cannot be found with the methods discussed in this paper.  One possible approach would be to scale out the $r^{-1/2}$ behavior explicitly, and solve only for the remaining parts of the solution, which should then be regular everywhere (compare \cite{HanHO09}).

\acknowledgments

It is a pleasure to thank Niall \'{O} Murchadha and Mark Hannam for very useful conversations, and Morgan MacLeod for his help with both computers and graphs.  JDI gratefully acknowledges support through a Maine Space Grant Consortium undergraduate student fellowship.  This work was supported in part by NSF grant  PHY-0756514 to Bowdoin College.

  
\end{document}